\documentclass[10pt,a4paper,twoside]{article}

\usepackage{vmargin,fancyheadings,multicol,multirow,ifthen,cite,
    epsfig,wrapfig,calc,dcolumn,apalike, graphicx,
    setspace, 
    boxedminipage,rotating,
    textcomp, 
        booktabs, 
    picinpar,longtable}

\usepackage{wgn2} 

\begin{document}
\SetPaperBodyFont   



\setcounter{page}{1}

\begin{WGNpaper}{twocol}{

\title{Comparison of three different camera systems monitoring the meteor activity over Hungary in 2020-2023}

\author{
L\'ivia Deme, Kriszti\'an S\'arneczky, Antal Igaz, Bal\'azs Cs\'ak, N\'andor Opitz, N\'ora Egei, J\'ozsef Vink\'o
(HUN-REN CSFK Konkoly Observatory)
\thanks{HUN-REN CSFK Konkoly Observatory, Konkoly Thege \'ut 15-17, Budapest, 1121 Hungary \\
        Email: {\tt vinko@konkoly.hu}}
}

\abstract{We present statistical analysis of visual meteor data taken with networks of meteor cameras operating in Hungary between 2020 and 2023. We use three different camera systems: a set of traditional MetRec-based video cameras, a self-developed automated DSLR camera system and a network of newly installed AllSky7 camera stations. Similarities and differences between the data produced by the three systems, aimed at recording different types of meteor phenomena, are presented and discussed.
}
\datereceived{Received 2023 December 19} 
}
\section{Introduction}
    \thankstext{HUN-REN Research Centre for Astronomy and Earth Sciences, Konkoly Observatory, Konkoly Thege \'ut 15-17, Budapest, 1121 Hungary \\
        Email: {\tt vinko@konkoly.hu}}
        
    \setcounter{footnote}{0} 

Dust grains  from cometary tails and/or ejected asteroid material, ranging from several microns to roughly meter-sized bodies in diameter, are responsible for most of the observed meteor events when entering Earth's atmosphere (see e.g. \cite{ye23} and references therein). Meteor populations are usually divided into two basic categories: sporadic meteors having no preferred direction, i.e. they appear at random positions on the sky all year round, and meteor showers arriving from a specific direction, called the radiant.  Meteoroid orbits causing many of the currently known meteor showers can be traced back to a parent body, usually a comet or an asteroid (\cite{mcintosh91,jen94,ye23}). Sporadic meteors are thought to be originated from old cometary tails and/or interplanetary dust trails that have been dissolved since their ejection, providing a more-or-less stationary background of incoming sporadic meteor flux. 

Since the visual meteor activity, i.e. the hourly or nightly rate of observable meteors are usually increased by more than an order of magnitude during the peak date of showers, meteor showers have been studied more frequently in the literature. Still, there is an increasing number of papers dealing with the statistical properties of sporadic meteors (see e.g. \cite{jb93}, \cite{rendtel06}, \cite{wiegert09}, \cite{dubi10} and references therein).

Despite the lack of a specific radiant, several preferred directions of sporadic meteors can be identified (e.g. \cite{cbj06}): most of the sporadic meteors arrive from the apex of Earth's orbit, while the  helion and antihelion sources are located in the ecliptic roughly perpendicular to the apex-antapex direction. Beside these sources, two additional ones at $\sim 60$ degrees north and south of the ecliptic, named northern and southern toroidal sources, were identified from radar observations (\cite{jb93}). The apex and the antihelion sources provide the majority of the visually observable sporaric meteor population (\cite{rendtel06}). 

In this paper we use data from different meteor camera systems operating in Hungary to reveal the visual meteor activity between 2020 and 2023. Here we concentrate on the statistical properties of sporadic meteors and the brightest events, mostly fireballs. The measured properties of recent meteor showers will be presented in a subsequent paper. In the next section we briefly summarize the technical parameters of the camera systems applied in this study, followed by the description of the data taken by the different cameras, the obtained results and our conclusions. 

\section{Meteor camera systems in Hungary}

In Hungary, the deployment of video camera systems for automated detection of visual meteors was started more than 20 years ago. Most of those cameras are still operating. They consist of an analogue video camera (either PAL, or NTSC) attached to a PC having a Matrox frame-grabber video card that samples the analogue video stream, and takes rapid snapshots of digital frames ($\sim 25$ frames per second) when a moving object, presumably a meteor, appears in the field-of-view. Meteor detection and measurement is handled by the "Meteor Recognizer" (MetRec, {\tt https://metrec.org}) software. The sensitivity limit is typically in between 4 and 5 magnitude, depending on the type and manufacturer of the particular camera. These cameras are integrated within the camera network of the International Meteor Organization (IMO). In this paper we use three of them: {\tt HUKON}, {\tt HUPIS} and {\tt HUHOD}. The basic parameters for these three cameras are summarized in Table~\ref{tab:metrec-camera-parameters}. MetRec uses a well-defined set of reference stars for computing astrometry and calibrated photometry for each meteor event. 

More recently, starting in 2017, our team at Konkoly Observatory developed and installed a new system, named Konkoly Meteor Observatory Network (KoMON), which is based on a combination of a digital video camera with a more modern DSLR digital camera. The moving object is recognized by the video camera uses a custom Python-based software then it triggers the DSLR camera to take digital frames for 10 seconds. An LCD panel, placed in front of the CMOS chip of the DSLR camera, chops the incoming light by a pre-programmed de Bruijn sequence (\cite{howie17}) to separate the likely meteors from other, slow-moving objects (airplanes, birds, treeleafs, insects, satellites, etc). The KoMON system has been designed to be sensitive to the brightest, relatively slow meteors, like fireballs. Such meteors might be missed by the conventional video camera systems, because they might saturate the camera causing a failure of the recognition algorithm. Thus, the KoMON system detects only the brightest end of the meteor brightness distribution function (BDF), and, by design, it likely detects the slower events instead of the fastest ones. The system parameters are collected in Table~\ref{tab:komon-camera-parameters}. 

At the time of writing this paper, proper astrometry and photometry is not yet implemented for the KoMON system. Thus, the cameras in the KoMON network are capable of recording only the moment (within $\sim$1 seconds) of the beginning of a meteor event and its angular speed along the projected trajectory. An integrated digital image (taken with 10 seconds exposure time) as well as digital video frames are also stored for each triggered event. Due to the lack of calibrated digital photometry, we were only able to use a visual brightness estimate for the recorded meteors that appeared brighter than the stars on the corresponding digital frame. This way we can still classify meteors that are brighter than $\sim 0$ visual magnitude, beside the lack of more precise photometric information. 

Finally, starting in mid-2021, several stations of the AllSky7 camera system ({\tt https://allsky7.net}) have been deployed at various sites in Hungary. These stations consist of an integrated unit of 7 cameras providing practically all-sky coverage, calibrated astrometry and broad brightness sensitivity range. The average limiting magnitude for each unit is $\sim 4$ mag. The parameters for the stations that are at our group's disposal are summarized in Table~\ref{tab:ams-camera-parameters}. 

\begin{table*}[]
    \centering
    \begin{tabular}{lcccccc}
         Name & Longitude & Latitude & Altitude & Center Az & Center Alt & Field-of-view \\
              & (deg.) & (deg.) & (m) & (deg.) & (deg.) & (deg. x deg.) \\
        \hline      
         HUKON & 18.963793 & 47.499622 & 490 & 317.7 & 40.9 & 80.7 x 56.4 \\
         HUPIS & 19.895140 & 47.917229 & 947 & 348.1 & 50.8 & 80.9 x 56.8 \\
         HUHOD & 20.312500 & 46.418499 & 79 & 328.7 & 50.8 & 80.3 x 55.8 \\
         \\
    \end{tabular}
    \caption{Parameters of the MetRec-based video cameras}
    \label{tab:metrec-camera-parameters}
\end{table*}

\begin{table*}
    \centering
    \begin{tabular}{lccccc}
    Site name & Longitude & Latitude & Altitude & Direction & Field-of-view \\
              & (deg.) & (deg.) & (m) &  & (deg. x deg.) \\
    \hline
    Konkoly & 18.963793 & 47.499622 & 490 & W & 90 x 80 \\
    Piszkesteto & 19.895140 & 47.917229 & 947 & Z,N,E,S,W & 360 x 90 \\
    Gothard & 16.6031 & 47.2578 & 232 & Z,N,E,S & 270 x 90 \\
    Devavanya & 20.9356 & 47.0814 & 84 & Z,N,E,S,W & 360 x 90 \\
    Becsehely & 16.7918 & 46.4479 & 180 & Z,N,E & 180 x 90 \\
    \\
    \end{tabular}
    \caption{Camera parameters for the KoMON stations}
    \label{tab:komon-camera-parameters}
\end{table*}

\begin{table*}
    \centering
    \begin{tabular}{lccccc}
    Site name & Camera name & Longitude & Latitude & Altitude & Field-of-view \\
      &  & (deg.) & (deg.) & (m) & \\
    \hline
    Konkoly & AMS18 & 18.963793 & 47.499622 & 490 & All-sky \\
    Becsehely & AMS71 & 16.7918 & 46.4479 & 180 & All-sky \\
    Piszkesteto & AMS72 & 19.895140 & 47.917229 & 947 & All-sky \\
    Fehergyarmat & AMS97 & 22.5178 & 47.9854 & 111 & All-sky \\
    Hortobagy & AMS98 & 21.1451 & 47.5941 & 121 & All-sky \\
    \\
    \end{tabular}
    \caption{Camera parameters for the AllSky7 stations}
    \label{tab:ams-camera-parameters}
\end{table*}

\section{Data}

The three different camera systems provided different types of data for us. The most complete high-level dataset was obtained by the MetRec-based cameras, after post-processing and cleaning the nightly data by visual inspection. MetRec is able to classify each recorded meteor based on its projected trail on the sky. If a meteor trail is close to a radiant of a known meteor shower, { and the meteor angular velocity matches to the known velocity of that meteor shower}, then the meteor is classified as a member of that particular shower. The sporadic meteors belonging to the antihelion source are classified as {\tt "ANT"}. Two other ecliptical showers, the Northern and Southern Taurids ({\tt NTA} and {\tt STA}, respectively) are also thought to belong to the sporadic meteor population. { All other meteors that cannot be assigned to any meteor showers are classified by MetRec as SPO. Many of those meteors are coming roughly from the apex direction}.  We extracted and analyzed the following data from the logfile of each night: date, time (both in UT), classification and magnitude for each event. We also used the nightly statistics computed by the auxiliary program {\tt CheckLog}: the total number of meteors, the number of sporadic meteors and the effective observing time for each recorded night. These numbers were then summed up for each month to create a monthly statistics from 2019 December to 2023 October. 

The KoMON system provided a more limited dataset. We used only the UT dates and time for each event recorded by each camera since 2019 December up to 2023 October. Simultaneous detections were treated as a single event, using a maximum time difference of $\pm 1$ second for multiple detections of the same meteor. We also visually inspected all frames to identify potential fireballs by selecting meteors that appeared brighter than any of the stellar objects on each frame (see Section 2). Finally, we  corrected the number of detected meteors for the number of active cameras at each site, resulting in a number of detections per a single DSLR camera FoV on each observed night. 

The AllSky7 system uses a custom Python-based software for meteor detection, astrometry and brightness estimates. Neither classification of meteor showers, nor calibrated photometry is done { at the individual sites, though. The meteor shower classification is part of the server-based multi-station analysis of the AllSky7 datasets.} Since the primary goal of the present study is meteor statistics, we extracted and used only the date and time of each event recorded by the AllSky7 stations at each site. 

\section{Results}

In this section we describe our results based on the data obtained by the three camera systems. 

\subsection{The population index of sporadic meteors}

Based on the magnitude information and classification for each meteors by MetRec, we were able to determine the nightly variation of the population index for the sporadic meteors, after summing up the number of events labeled as {\tt SPO}, {\tt ANT}, {\tt NTA} or {\tt STA}. 

The population index is determined from the observed cumulative BDF for a sample of meteors: 
\begin{equation}
    N(m) = N_0 \cdot r^{m} ~~~ (m < m^*), 
    \label{eq:popindex}
\end{equation}
where $m$ is the magnitude, $N(m)$ is the number of meteors brighter than $m$ magnitude, $N_0$ is a normalization constant,  $m^*$ is the magnitude limit for detection completeness and $r$ is the population index. From Equation~(\ref{eq:popindex}) the population index can be derived simply as $r = N(m+1) / N(m)$ for $m < m^*$. 

We adopted $m^*$ as the magnitude of the peak of the meteor BDF histogram using 1 magnitude-wide bins, and derived the population index using the numbers for 2 magnitude bins below the peak. Figure~\ref{fig:popindex1} illustrates the methodology: the left panel shows the histogram of the BDF of all sporadic meteors from the {\tt HUKON} camera, { which turned out to be the most stable one from photometric point of view.} The right panel plots the cumulative BDF for the same dataset. $m^* = 1$ is estimated as the magnitude of the maximum of the BDF in the left panel. The blue line is a least-squares fit of Equation~(\ref{eq:popindex}) to the points within the interval of $[m^* - 2 < m < m^*]$. 

\begin{figure}
    \centering
    \includegraphics[width=8cm,height=5cm]{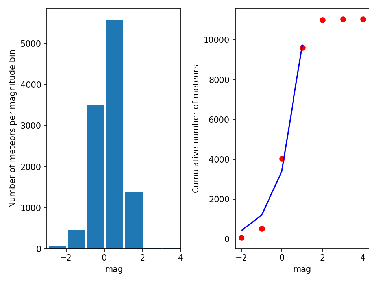}
    \caption{The method used for population index determination. Left panel: the observed number of meteors ($n(m)$)  as a function of magnitude. Right panel: the cumulative BDF ($N(m)$) against magnitude. The blue line indicates the fitted model from Eq.~\ref{eq:popindex}.}
    \label{fig:popindex1}
\end{figure}

Figure~\ref{fig:popindex} shows the variation of the nightly population indices for the three {\tt MetRec} cameras as a function of Julian dates. Ignoring the few outliers (probably due to low number of recorded meteors and/or inferior weather conditons), most of the indices are within the range of { $2 < r < 4$}. The black horizontal line indicates the mean value of { $<r> = 2.85 \pm 0.48$} for the combined sample of the three cameras.
This is { in good agreement with} result of \cite{rendtel06} who obtained $<r> = 2.95 \pm 0.15$ as a yearly average. 

Note, however, that the population index determination suffers from known issues. For example, \cite{molau15} discussed several improvements which may result in somewhat different population indices for the same dataset. He obtained $<r> \approx 2.5$ as a mean value for sporadic meteors, 
which is very similar to the results of \cite{jen94} (2.2 - 2.5) and \cite{vida20} ($<r> = 2.55 \pm 0.06$). 
More recently \cite{betzler23} suggested a different statistical function (the q-exponential function) to represent the cumulative meteor BDF. He obtained $<r> = 3.63 \pm 0.01$ based on a more extended magnitude range of $-5 < m < 0$. The differences between these reported population indices illustrate that despite the improvement of the methodology, these estimates may still suffer some sort of systematic errors. Nevertheless, in the following we apply our results above, { $<r> = 2.85 \pm 0.48$} for consistency. Contrary to \cite{betzler23}, we did not detect significant variation in the population index of sporadic meteors along a year (Figure~\ref{fig:popindex}).

\begin{figure}
    \centering
    \includegraphics[width=8cm]{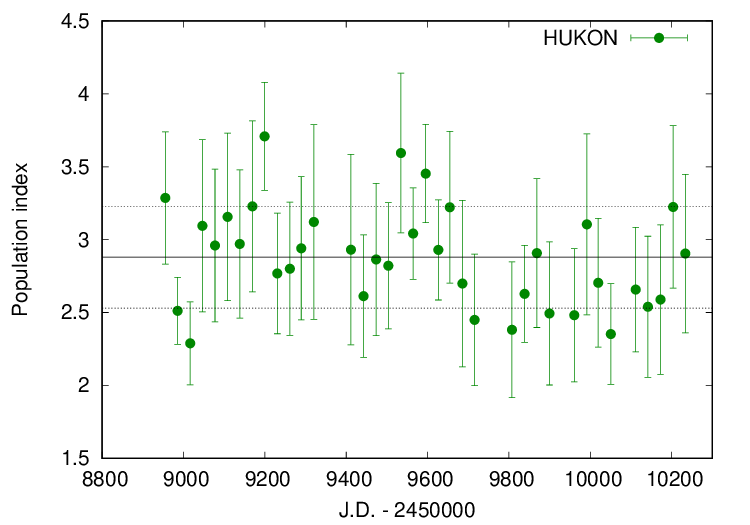}
    \caption{The variation of the nightly population index of {\tt SPO} meteors recorded by {\tt HUKON} during 2020-2023.}
    \label{fig:popindex}
\end{figure}

\subsection{The hourly rate of sporadic meteors}

\begin{figure*}
    \centering
    \includegraphics[width=8cm,height=7cm]{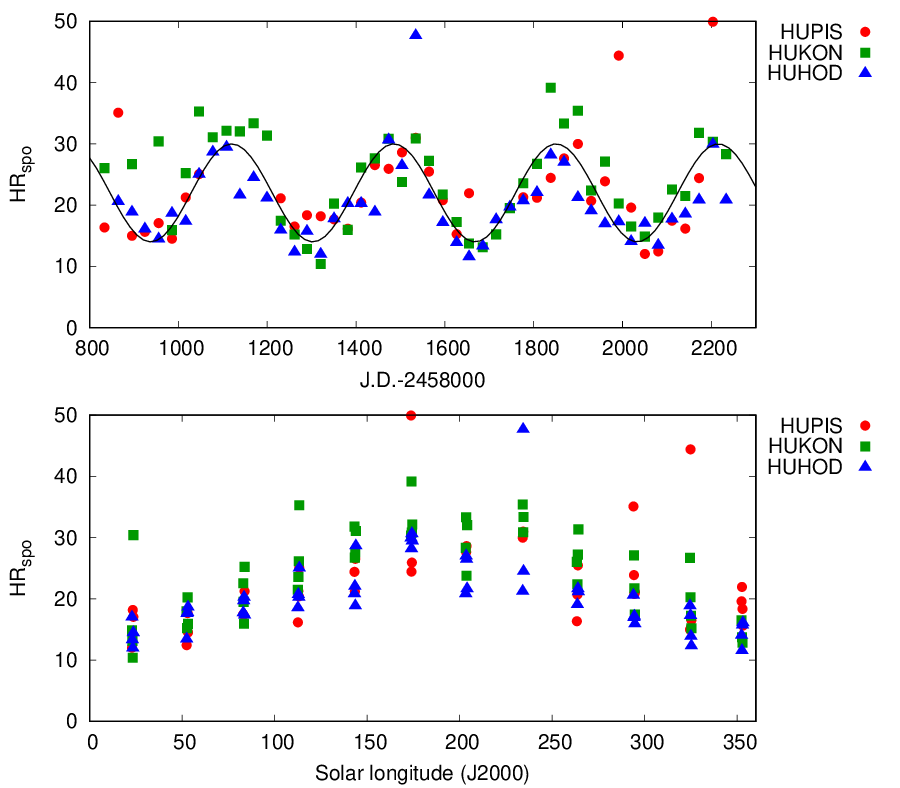}
    \includegraphics[width=8cm,height=7.5cm]{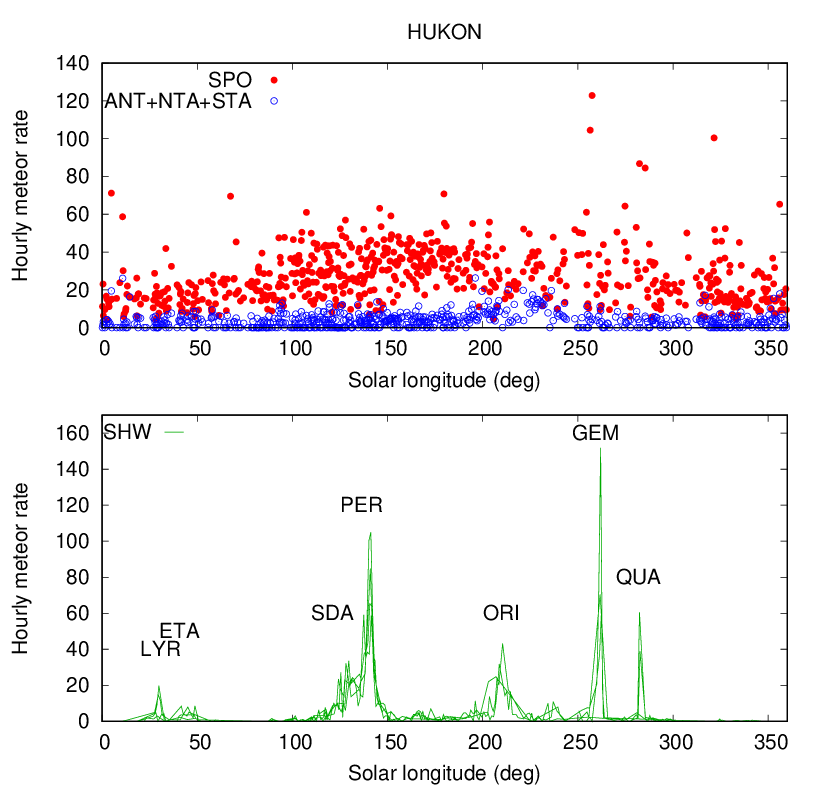}
    \caption{Left: The hourly rates of {\tt SPO} meteors as a function of J.D. (upper panel) and solar longitude (lower panel). Right: Hourly meteor rates by HUKON for all sporadic meteors, including {\tt SPO, ANT, NTA} and {\tt STA} (upper panel) and meteor showers (lower panel). Major showers are labeled by their IMO codes. }
    \label{fig:metrec_rates_new}
\end{figure*}

The rates of sporadic meteors were estimated from the {\tt HUKON}, {\tt HUPIS} and {\tt HUHOD} data in the following way. We adopt the sporadic hourly rate ($\mathrm{HR_{spo}}$) definition from \cite{dubi10} as
\begin{equation}
\mathrm{HR_{spo}} ~=~ \frac{n_{spo}}{t_{eff}} \cdot F \cdot r^{\Delta m},
\label{eq:hrspo}
\end{equation}
where $n_{spo}$ is the number of sporadic meteors per night, $r$ is the population index, $\Delta m$ is the magnitude difference between the camera sensitivity limit and a reference value, $t_{eff}$ is the effective observation time in hours and $F$ is the correction factor { between the camera field-of-view (FoV) and that of a human observer. For the latter we adopt a conservative estimate that a human observer sees $\sim 1/3$ of the visible hemisphere of the sky.}  From Equation~\ref{eq:hrspo} it is seen that we do not correct the rate for the zenith distance of a fiducial "radiant" of sporadic meteors, unlike e.g. \cite{rendtel06}. It is a common practice to define $\Delta m = 6.5 - Lm$, where $Lm$ is the limiting magnitude (the sensitivity limit) for a particular camera (\cite{jen94}). 

{ We adopted the following nominal limiting magnitudes: $Lm = 4.5$ ({\tt HUKON} and {\tt HUPIS}) and $Lm = 4.0$ ({\tt HUHOD}). Since these limiting magnitudes may change during the night due to variable weather conditions, we also attempted to use the nightly averaged values from the "MAG" files provided by MetRec. The limiting magnitude estimates by MetRec have a 1 minute resolution, thus, they could be used to model the variable weather conditions during each night. However, it was found that while using a variable $\Delta m$ based on these data the night-to-night scattering of the calculated $\mathrm{HR_{spo}}$ values increases by a large amount, which is not physically expected for the true rate of sporadic meteors. Therefore, we decided to apply a single $\Delta m$ value for each camera as listed above. }

Figure~\ref{fig:metrec_rates_new} shows the results. In the upper left panel we plot $\mathrm{HR_{spo}}$ for all three cameras 
against Julian days. The annual oscillation, noted previously by many observers (e.g. \cite{cbj06}; \cite{rendtel06}) is evident from these data. \cite{dubi10} even found correlation between $\mathrm{HR_{spo}}$ and the solar activity cycle. Our data do not cover $\sim 20$ years, which would be necessary to confirm the existence of a $\sim 11$ year-long cycle, but the annual variation is recovered consistently by all three cameras. 

The lower left panel of Figure~\ref{fig:metrec_rates_new} displays the same data against solar longitude. Again, the increase of $\mathrm{HR_{spo}}$ during the Fall season (September - November) is visible each year. This is fully consistent with the results of \cite{dubi10} who also found a $\sim 50$ percent increase of $\mathrm{HR_{spo}}$ in September with respect to the rate measured in April. { This annual variation of $\mathrm{HR_{spo}}$ seems to be consistent with the seasonal visibility of the apex direction during the 2nd half of the night.}

\cite{rendtel06} attributed the annual variation of $\mathrm{HR_{spo}}$ he found in the IMO database (VMDB) to the polluting effect of misclassified shower meteors (see Figure 6 in \cite{rendtel06}). We also tested this effect, but found that this is less significant in our data. In the upper right panel of Figure~\ref{fig:metrec_rates_new} we plot the nightly $\mathrm{HR_{spo}}$ derived from the {\tt HUKON} data as a function of solar longitude. 
It is seen that the {\tt SPO} meteors (plotted with red symbols) dominate the majority of sporadic events, and the increase of the sporadic activity within $\lambda = 100$ - 200 deg is mostly due to {\tt SPO} meteors. The antihelion source (plotted with blue symbols) provides only a small fraction of the total sporadic rate, and the annual variation is less visible in those data. 

The lower right panel displays the inferred rates for meteor showers. These are only approximate ZHRs as they are not corrected for radiant position to keep consistency with the measurement of the sporadic rates, and they are used only for comparison. Indeed, the appearance of several of the richest showers can be easily identified, and they are labeled. Comparing the top right panel with the bottom right figure, it is seen that the increase of the sporadic rate is not generally due to the polluting effect of showers. For example, in the late August - late October season ($\lambda = 150$ - 200) , when $\mathrm{HR_{spo}}$ is the highest in our data, the activity from showers was negligible, at least from 2020 to 2023. On the other hand, a pollution effect can be identified during the Geminids in mid-December, but this is localized to a single night during the shower maximum, around December 13. Thus, we conclude that the increase of $\mathrm{HR_{spo}}$ during September - November is real. { It correlates well with the seasonal visibility of the apex direction where most of the sporadic meteors are coming from, but it might also be} connected with some sort of enhancement in the distributions of meteoroid parameters (mass, particle size, velocity, spatial density, etc) that contribute to the appearance of visually identified meteors, observed locally from the northern hemisphere (\cite{dubi10}). 

Regarding the median value of $\mathrm{HR_{spo}}$ and its standard deviation, we measure $26.3 \pm 6.8$, $32.1 \pm 9.2$ and $26.9 \pm 9.1$ for {\tt HUPIS}, {\tt HUKON} and {\tt HUHOD}, respectively. For the combined sample it is $28.4 \pm 8.4$. 
{ These values are generally higher than those in recent literature: they are a factor of $\sim 3$ higher than the results of \cite{dubi10} ($\mathrm{HR_{spo}} \sim 10 \pm 2$ from data by visual observers), and still somewhat higher than the estimate by \cite{rendtel06} based on video meteor observations ($\mathrm{HR_{spo}} \sim 22 \pm 3$). Since $\mathrm{HR_{spo}}$ as defined in Equation~\ref{eq:hrspo} depends critically on the assumed FoV of a visual observer as well as the proper statistical accounting of the unobserved faint meteors, the difference between our results and that of \cite{rendtel06} is not significant. Note that if we had corrected the camera FoV to the whole area of the full sky hemishere ($2 \pi$ steradian) instead of the FoV of a human observer, then the calculated rates should have been scaled up to 3 times higher values, resulting in $\sim 85$ sporadic meteors/hour. 
}

\subsection{The frequency of slow fireballs as potential impactors}

\begin{figure}
    \centering
    \includegraphics[width=8cm,height=7cm]{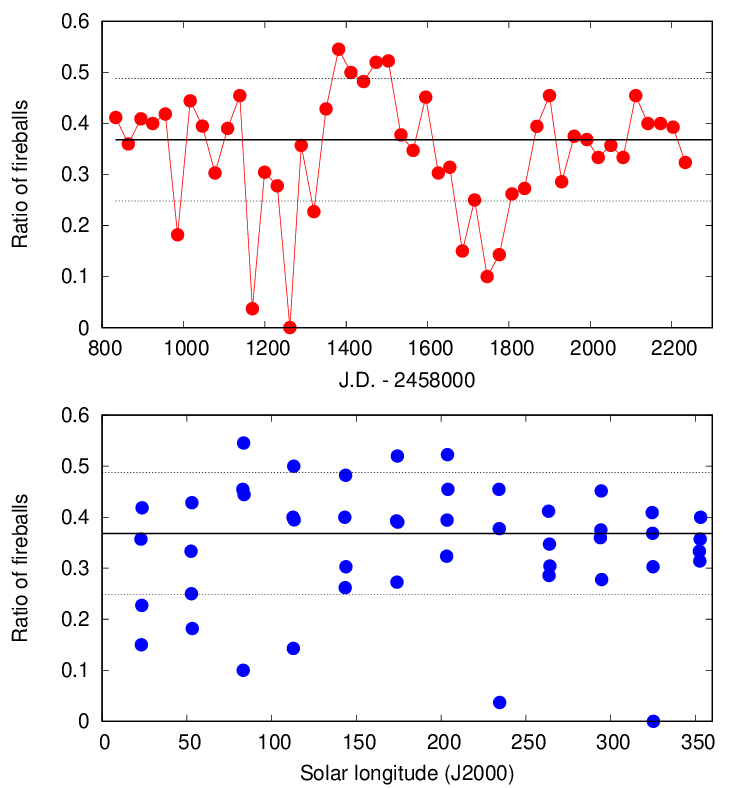}
    \caption{The ratio of "fireballs" (meteors brighter than 0 mag in this paper) to the number of all meteors recorded by the KoMON system as a function of J.D. (top panel) and solar longitude (bottom panel)}
    \label{fig:fireball-ratio}
\end{figure}

Fireballs are spectacular events representing the brightest end of the BDF of visual meteors. Traditionally, meteors brighter than Venus ($m < -4$ mag) are called fireballs. Most of them produce { flares} during or at the end of their visible path, but the definition is related only to the visual brightness (i.e. brighter than $-4$ magnitude) and not the occurrence of any { flares}. Since fireballs are the most probable impactors among meteors, there is significant interest in studying their frequency in any meteor stream. 

The KoMON system, as mentioned in Section~2, was designed to capture mostly the brightest and relatively slow meteors, in order to detect potential impactors. Even though its limited capabilities, we were able to visually identify the brightest events (brighter than $\sim 0$ magnitude) by visually inspecting every recorded frame, starting from 2019 December up to 2023 October. The reason for extending the magnitude limit to 0 mag, instead of $-4$ mag, was twofold. First, since there is no calibrated photometry available in the KoMON system, only relative brightness estimates with respect to the brightest visible stars could be made. Second, since the real fireballs are rare, extending the magnitude limit from $-4$ mag down to 0 mag increases the number of events in the sample, thus, it becomes more suitable for statistical studies. In the following, when we refer to fireballs, we mean meteors brighter than 0 magnitude. 

Figure~\ref{fig:fireball-ratio} plots the ratio of fireballs to all events recorded by the KoMON cameras as a function of Julian dates (upper panel) and solar longitude (lower panel). 
Meteor numbers belonging to the same month are summed, in order to get better statistics for these relatively rare events. 
Thus, each symbol represent the ratio of the monthly sum of fireballs to the sum of all recorded meteors. As mentioned in Section~2, multiple detections have been removed, thus, they do not introduce a bias in the number of fireballs. The mean fireball ratio for the whole sample turned out to be about 30 - 50 percent ($0.37 \pm 0.12$). The actual numbers go below or above this range in a few months when very few meteors were detected by the KoMON cameras for a number of reasons (weather, technical issues, etc). 

\begin{figure}
    \centering
    \includegraphics[width=8cm,height=7cm]{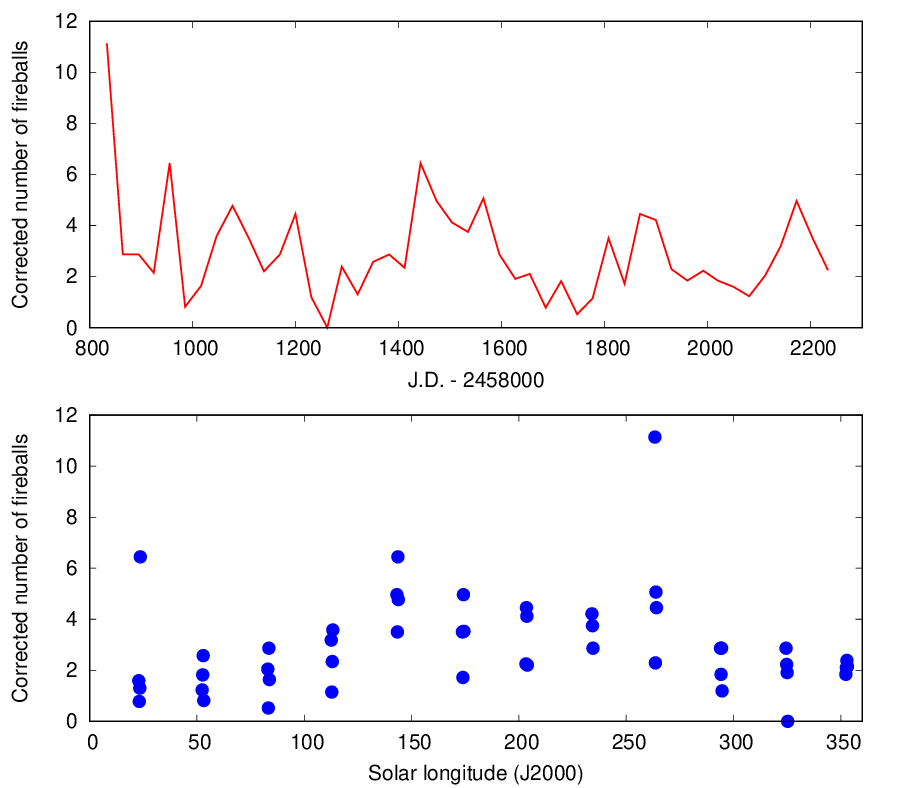}
    \caption{The number of fireball events per month scaled up to all-sky as a function of J.D. (upper panel) and solar longitude (lower panel). The Geminids and Perseids are clearly identifiable as local maxima. }
    \label{fig:meteor_stat_fireball}
\end{figure}

Figure~\ref{fig:meteor_stat_fireball} displays the number of fireballs per month, normalized to a single camera and scaled up to all-sky FoV, again, as a function of J.D. (upper panel) and solar longitude (lower panel). These numbers can be regarded as the number of fireballs in each month for the whole sky above Hungary. Again, we used the monthly sums to reduce the fluctuations due to low number statistics. As expected, meteor showers are clearly identifiable in these plots. For example, Geminids in 2019 December produced  especially high number of fireballs. Later Perseids at $\lambda \sim 145$ deg showed the highest number of fireballs recorded by the KoMON system. The mean value is estimated as $N_{fb} = 2.94 \pm 1.91$ fireballs per month. 

\begin{figure}
    \centering
    \includegraphics[width=8cm,height=7cm]{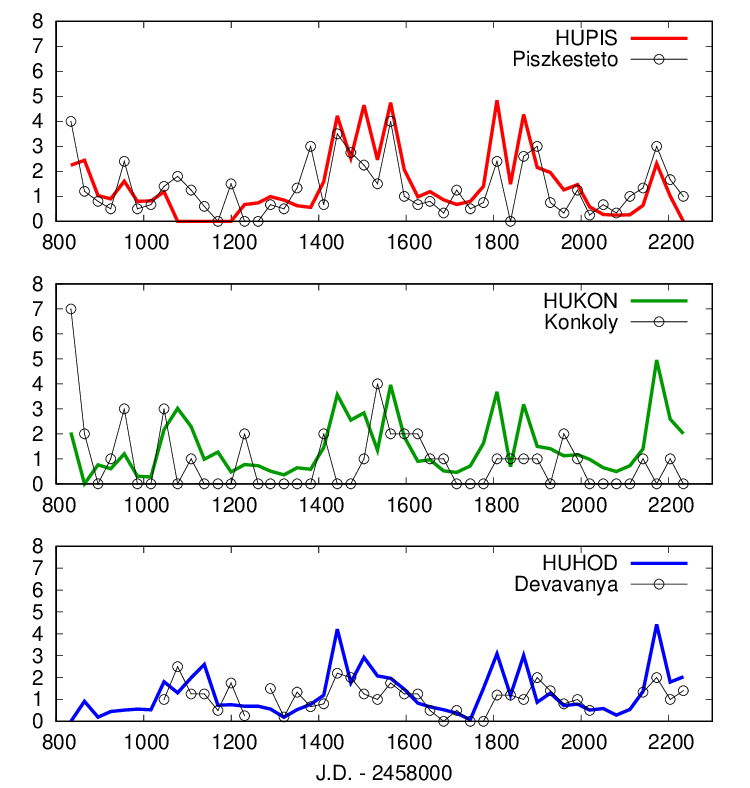}
    \caption{The comparison of the KoMON fireball numbers normalized to a single camera (black lines with dots) and the MetRec monthly meteor numbers corrected for 0 magnitude (color curves). See text for explanation.}
    \label{fig:fireball_predict}
\end{figure}

Finally, we compare the KoMON fireball numbers to the meteor numbers taken by the MetRec-based cameras after statistically selecting only the brightest meteors. Figure~\ref{fig:fireball_predict} shows the comparison for those cameras that are either at the same site (Piszkesteto - {\tt HUPIS}, Konkoly - {\tt HUKON}) or relatively nearby (Devavanya - {\tt HUHOD}). Since the camera FoV-s in the two systems are similar { ($81^o$x$56^o$ for a MetRec camera and $90^o$x$80^o$ for a single KoMON camera)}, no correction for this factor was made. The KoMON data taken with multiple units at the same site were normalized to a single camera. 
The numbers from the MetRec-based cameras are corrected for the magnitude difference between the camera limiting magnitude and 0 mag, thus, applying the factor of $r^{\Delta m}$, where $r$ is the population index. We found that if we set $r = 3.0$ and $\Delta m = -5.0$, $-5.0$ and $-4.5$ for {\tt HUPIS}, {\tt HUKON} and {\tt HUHOD}, respectively, they give the best agreement between the MetRec and the KoMON data. It is seen in Figure~\ref{fig:fireball_predict} that the two datasets are generally consistent. Thus, the KoMON system is able to capture most of the meteors at the brightest tail of the meteor BDF if they appear within the camera field-of-view. 

\begin{figure}
    \centering
    \includegraphics[width=8cm]{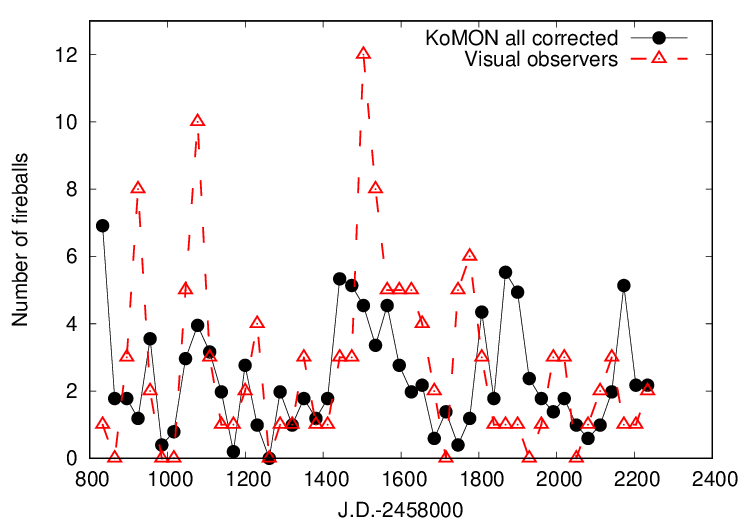}
    \caption{The total number of KoMON fireballs corrected for the $-4$ magnitude limit (assuming a population index of $r=1.5$, see text) plotted with filled red circles and the number of observed fireballs brighter than $-4$ magnitude reported by visual observers from Hungary (green open triangles). }
    \label{fig:fireball_vis}
\end{figure}

It is interesting how many true fireballs we can see if we use the original $-4$ mag limit as a cutoff. Using the same population index of $r = 3.0$, the normalized KoMON fireball numbers must be corrected with $r^{-4} \approx 0.012$. Such a correction would predict an average of $\sim 0.036$ fireball per month, i.e. a single fireball in every 2 years. This is clearly way below the actual number of detected fireballs, either by our cameras or by visual observers. { Thus, the population index may be different from 3. This is expected because most of the true fireballs are due to shower meteors that have lower population index. Assuming $r = 1.5$ one would get the true fireball rate as $\sim 0.6$ per camera per month, which is a more reasonable estimate. }

{ We have made an independent test in order to confirm this assumption by comparing the predicted number of KoMON fireballs brighter than $-4$ magnitude using the correction factor of $r^{-4}$ with $r=1.5$ with the number of actually observed fireballs brighter than $-4$ magnitude over Hungary. For the former we selected all fireballs detected by the KoMON system, while for the latter data we used the Hungarian mirror of the AMS fireball report webpage ({\tt https://mcse.imo.net}) to collect the observed fireballs over Hungary as reported by visual observers between 2020 and 2023. Multiple detections were filtered out, so each fireball was counted only once. The results are plotted together in Figure~\ref{fig:fireball_vis}. It is seen that the two independent datasets are consistent with each other. It confirms the hypothesis that the magnitude distribution of bright fireballs can be modeled with the population index of $r \sim 1.5$, which is significantly less than that of the sporadic meteors. Also, based on the detections by the KoMON system one can predict the true number of very bright fireballs reasonably well.  
}

\subsection{Comparison of MetRec-based cameras with AllSky7 units}

\begin{figure*}
    \centering
    \includegraphics[width=8cm, height=5cm]{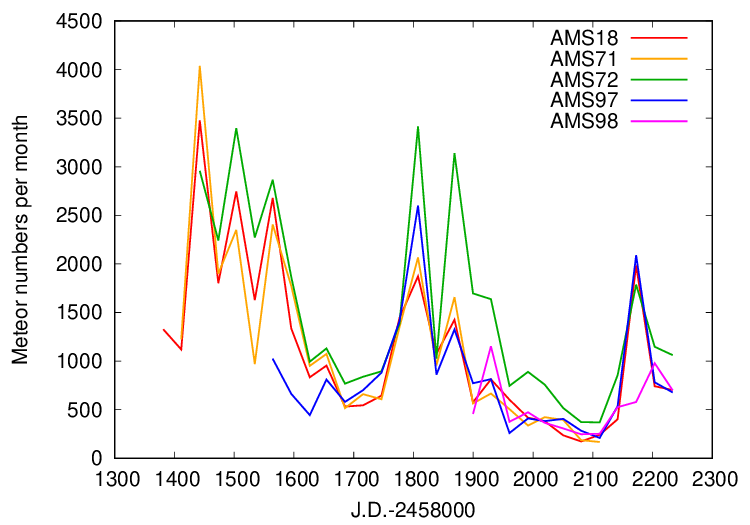}
    \includegraphics[width=8cm,height=5cm]{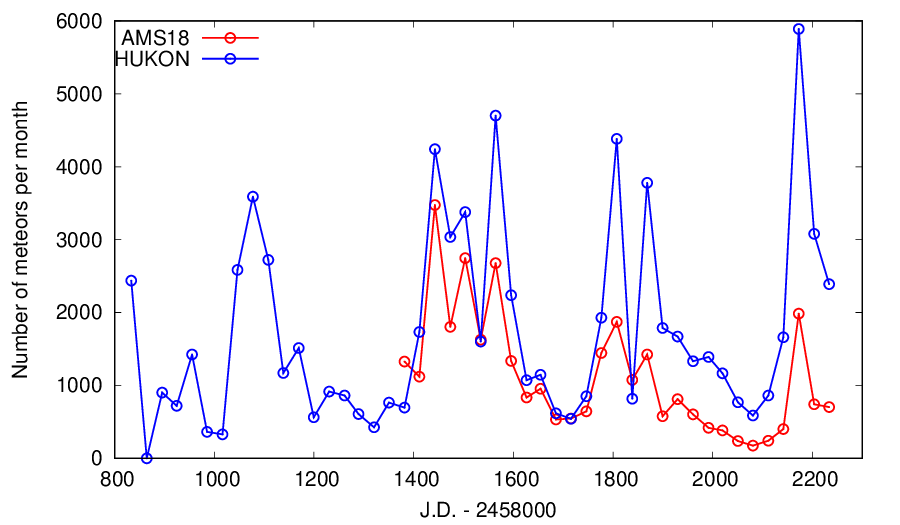}
    \caption{Left panel: monthly meteor detections for the AllSky7 cameras. Right panel: comparison of monthly meteor detections by AMS18 and HUKON, after correction for field-of-view difference.}
    \label{fig:ams_comp}
\end{figure*}

\begin{figure}
    \centering
    \includegraphics[width=8cm]{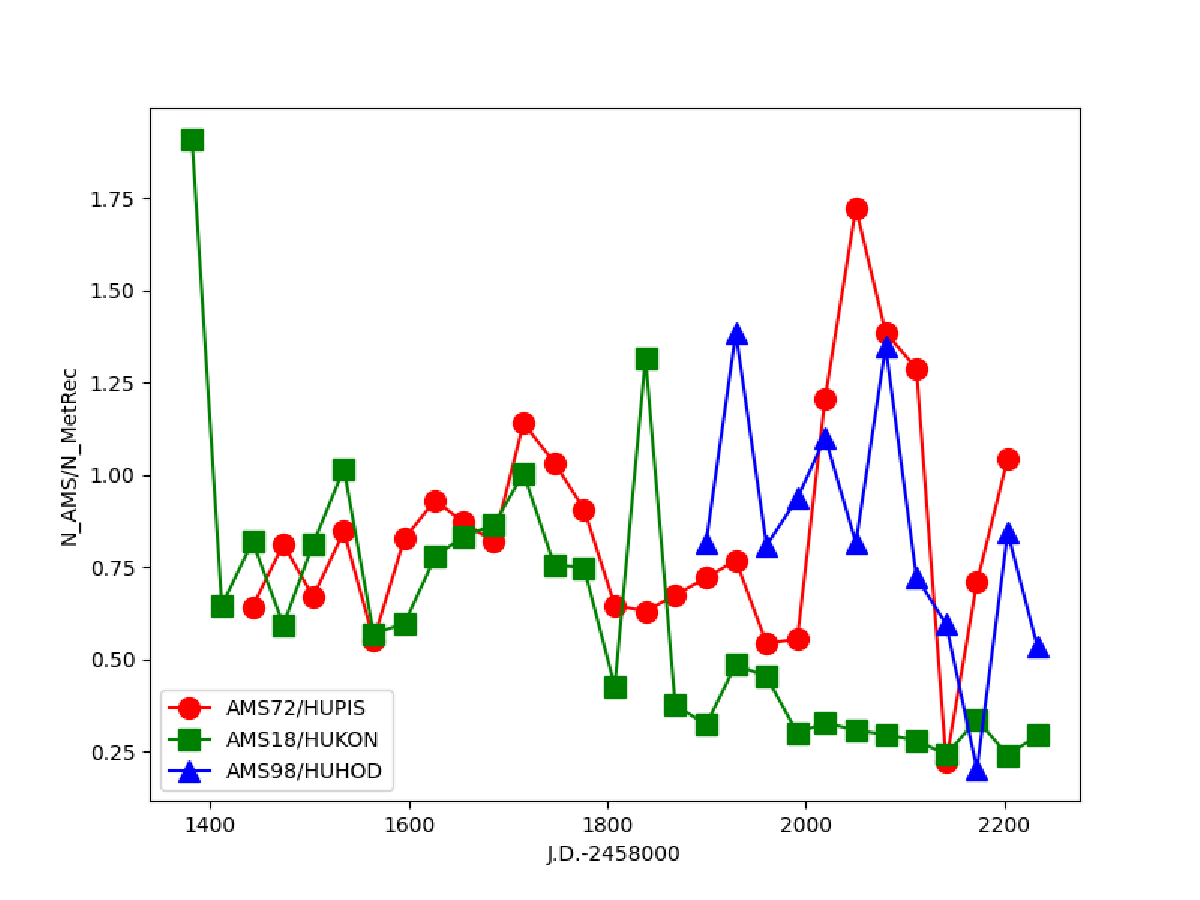}
    \caption{Ratio of AMS and MetRec meteor numbers.}
    \label{fig:ams_ratio}
\end{figure}

Since the AllSky7 units provide all-sky coverage, limiting magnitude comparable to that of the MetRec-based cameras and non-stop operation, they are expected to become the next generation of autonomous meteor cameras when the MetRec-based video cameras are retired. Thus, it is useful to compare the performance of the two camera systems in order to test whether the AllSky7 cameras delivered datasets similar to MetRec, both in quantity and quality. 

Because the Python-based software of the AllSky7 cameras are still under continuous development (Hankey, personal communication), in this paper we use only the total number of meteors recorded by the AllSky7 stations at our disposal (Table~\ref{tab:ams-camera-parameters}) as a test parameter. The meteor numbers were collected from the camera servers by summing up the number of "reduced" meteor events on each night, after visually inspecting all frames and removing the likely artificial records (usually due to airplanes).

In the left panel of Figure~\ref{fig:ams_comp} we plot the sum of the recorded meteors in each month since the deployment of the individual stations as a function of Julian date. It is seen at first glance that the summed meteor numbers from the different stations are more-or-less consistent with each other, even though the individual stations are distributed at $\sim 100$ - 200 km from each other, thus, they may experience different observing/weather conditions. On the other hand, it is also visible that the monthly meteor numbers show a declining trend for all cameras during the first $\sim 800$ days after deployment. This decline of the recorded meteor numbers seems to be less pronounced for AMS72 during the 1st year of its operation, but then it also develops the same declining trend in its recorded meteors. 

In the right panel of Figure~\ref{fig:ams_comp} the data from AMS18 are compared with the monthly sum of meteors captured by {\tt HUKON}. These two cameras are both placed at the same site, on the rooftop of the headquarters of Konkoly Observatory in Budapest, Hungary, thus, weather-related differences are not expected. The {\tt HUKON} data were corrected for the smaller field-of-view of the camera, thus, they represent the expected number of meteors as if {\tt HUKON} were an all-sky camera, like AMS18. 
It is seen that the mean level of the {\tt HUKON} data is relatively constant each year (except, of course, the peak amplitudes due to the different activity levels of the richest meteor showers). On the contrary, the meteor numbers from AMS18 show a notable decline with respect to the {\tt HUKON} data. Since these two cameras are at the same site, this illustrates that the decline is not due to the general and continuous worsening of the observing conditions (e.g. climate change) over Hungary. Instead, it is suspected that this effect is caused by some kind of technical issues affecting the AllSky7 camera stations. 

In Figure~\ref{fig:ams_ratio} we plot the ratio of the monthly meteor numbers from three different AllSky7 station and the nearest MetRec-based cameras. The decline in the AMS18/{\tt HUKON} data is clearly visible, as above. The AMS72/{\tt HUPIS} ratio seems to be more stable, but the second part of the curve (the data taken in 2023) shows wide oscillations preventing the observability of any long-term trend. The AMS98/{\tt HUHOD} ratio spans a shorter range than the other two, but it suggests the same declining trend as the AMS18 data. 

Thus, it is concluded that all AllSky7 stations (except maybe AMS72) managed by our group show a general long-term decline in the recorded monthly meteor numbers, while such a decline is not visible in the data of our MetRec-based cameras. { According to the developer of the AllSky7 system (Hankey, personal communication), this seems to be a software-related issue due to multiple reasons. One is related to the applied screening criterion: previously the software was designed to select meteors whose tracks contain at least three measured points, while later it was changed to keep only those that have at least 5 points. Also, sometimes the screening AI unintentionally misses some of the faint and short meteors. Since the AllSky7 system was specifically designed to be triggered by meteors that can be simultaneously measured from many sites, these changes resulted in better performance of the system as a whole. However, our analysis above shows that with the present conditions the data from the individual AllSky7 stations are less suitable for statistical studies than those from the MetRec-based cameras.}

\section{Conclusion}

In this paper we presented a statistical study of the data from three different meteor camera system operating in Hungary from 2020 to 2023. Based on the results detailed in the previous sections, we draw the following conclusions. 

\begin{enumerate}

\item{We measured the population index of all sporadic meteors detected by our MetRec-based cameras, and obtained $<r> = 2.85 \pm 0.48$ as the mean value and standard deviation. This is consistent with the results by \cite{jen94}, \cite{rendtel06} and \cite{vida20}. Unlike \cite{betzler23}, we did not detect annual variation in the population index of sporadic meteors.}

\item{The hourly rate of sporadic meteors was estimated based on the data provided by MetRec. We got $\mathrm{HR_{spo}} = 28.4 \pm 8.4$ sporadic meteors/hour for the combined sample from all three cameras ({\tt HUPIS}, {\tt HUKON} and {\tt HUHOD}). { This is consistent with the results of \cite{rendtel06} based on video meteor data, but seems to be a factor of 3 higher than the estimates by \cite{dubi10} from visual observations.} }

\item{The frequency of "fireballs" (defined as meteors brighter than 0 mag in this paper for statistical reasons) was studied by the KoMON camera system. We obtained a monthly rate of $N_{fb} \approx 2.94 \pm 1.91$ events per month. Comparing these data with those from the MetRec-based cameras after scaling those down to 0 limiting magnitude we got reasonable, order-of-magnitude agreement. Note that after applying the brightness correction to $-4$ mag using a population index of $r \sim 1.5$, we found { good agreement with the numbers of true fireballs reported by visual observers from Hungary. This is consistent with the expectations, because most fireballs likely belong to meteor showers, whose population index is substantially lower than that of the sporadic meteors.}  }

\item{Comparing the monthly total meteor numbers taken by our AllSky7 stations with similar data from MetRec (after scaling the latter to all-sky) we found an unexpected decrease of the AllSky7 data with respect to those from MetRec. All our AllSky7 stations show the same trend, except maybe AMS72. We tested that it is not related to any long-term worsening of the observing conditions, and it is not likely due to the varying contamination of non-meteor events in the AllSky7 data. { All the data suggest that it is probably a software-related reason, which has been confirmed by the developer.} }

\end{enumerate}

Acknowledgements:
This work was supported by the project "Cosmic Effects and Risks" GINOP 2.3.2-15-2016-0003 by the Hungarian National Research Development and Innovation Office, based on funding provided by the European Union. \\
Special thanks are due to Nagykanizsa Amateur Astronomers Association (especially Zsolt Perkó and Attila Gazdag), and the Bárdos Lajos Primary School, Fehérgyarmat (especially Zoltán Pásztor) for their kind contributions.

\bibliography{meteor.bib}



\end{WGNpaper}

\end{document}